\documentclass[letter]{aa}

\usepackage{natbib}
\bibpunct{(}{)}{;}{a}{}{,} 
\usepackage{graphics}   
\usepackage{graphicx}   
\usepackage{amsmath}
\usepackage{epstopdf}
\usepackage[varg]{txfonts}
\usepackage{pdflscape}
\usepackage{hyperref}   
\usepackage[flushleft]{threeparttable}

\hypersetup{
    bookmarks=true,         
    unicode=false,          
    colorlinks=true,       
    linkcolor=blue,          
    citecolor=blue,        
    filecolor=blue,      
    urlcolor=blue           
}

\begin{document} 

   \title{Are some CEMP-s stars the daughters of spinstars?}

   \author{
   Arthur Choplin
          \inst{1},
          Raphael Hirschi
          \inst{2,3,4},
           Georges Meynet
           \inst{1},
          \and
          Sylvia Ekstr\"{o}m
          \inst{1}
                    }

 \authorrunning{Choplin et al.}

        \institute{Geneva Observatory, University of Geneva, Maillettes 51, CH-1290 Sauverny, Switzerland\\
        e-mail: arthur.choplin@unige.ch
        \and Astrophysics Group, Lennard-Jones Labs 2.09, Keele University, ST5 5BG, Staffordshire, UK
        \and Kavli Institute for the Physics and Mathematics of the Universe (WPI), University of Tokyo, 5-1-5
Kashiwanoha, Kashiwa, 277-8583, Japan
        \and UK Network for Bridging the Disciplines of Galactic Chemical Evolution (BRIDGCE)
                }
   \date{Received / Accepted}
 
  \abstract
   {
   CEMP-s stars are long-lived low-mass stars with a very low iron content as well as overabundances of carbon and s-elements.
   Their peculiar chemical pattern is often explained by pollution from a AGB star companion.
   Recent observations have shown that most of the CEMP-s stars are in a binary system, providing support to the AGB companion scenario.
   A few CEMP-s stars, however, appear to be single.
   We inspect four apparently single CEMP-s stars and discuss the possibility that they formed from the ejecta of a previous-generation massive star, referred to as the ''source'' star.
   In order to investigate this scenario, we computed low-metallicity massive star models with and without rotation and including complete s-process nucleosynthesis.
   We find that non-rotating source stars cannot explain the observed abundance of any of the four CEMP-s stars.
   Three out of the four CEMP-s stars can be explained by a $25$ $M_{\odot}$ source star with $v_{\rm ini} \sim 500$ km s$^{-1}$ (spinstar). 
   The fourth CEMP-s star has a high Pb abundance that cannot be explained by any of the models we computed. Since spinstars and AGB predict different ranges of [O/Fe] and [ls/hs], these ratios could be an interesting way to further test these two scenarios.
   }

   \maketitle

        \titlerunning{Spinstars and CEMP-no stars}
        \authorrunning{A. Choplin et al.}

\section{Introduction}\label{sec:1}

Carbon-Enhanced Metal-Poor (CEMP) stars are iron-deficient stars with an excess of carbon compared to normal metal-poor stars. 
Some of these stars with very little iron, e.g. SMSS J031300.36-670839.3, with [Fe/H] $<-7$ \citep{keller14}, should have formed from a material ejected by the first massive stars in the Universe. 
Nowadays, CEMP stars are generally considered as the best window into the first stars. 

CEMP stars are divided in several subclasses depending on their enrichment in s- and r-elements: CEMP-s, CEMP-r, CEMP-r/s \citep[or CEMP-i,][]{hampel16} and CEMP-no (little enriched in s- and r-elements). 
Most CEMP-s stars have $-3<$ [Fe/H] $<-2$ \citep{norris13}.
Different scenarios are needed to explain CEMP stars \citep[even in CEMP subclasses, different classes of progenitors seem to be needed, e.g. for the CEMP-no category,][]{placco16, yoon16, choplin17}.
For CEMP-s stars, the main formation scenario is the AGB scenario, suggesting that a more massive AGB companion has fed the secondary in carbon and s-elements during a mass transfer (or wind mass transfer) episode \citep{stancliffe08, lau09,bisterzo10,lugaro12,abate13,abate15a,abate15b,hollek15}. Interestingly, it has been shown by \cite{matrozis17} that rotational mixing in the CEMP-s stars can severely inhibit atomic diffusion. If not counteracted, atomic diffusion would make the s-elements sink quickly into the star after the accretion episode.
By considering a sample of CEMP-s stars, \cite{lucatello05} and \cite{starkenburg14} have shown that the whole sample is consistent with the hypothesis of all being in a binary system. \cite{hansen16a} have monitored the radial velocity of 22 CEMP-s stars during several years. They found clear orbital motion for 18 stars, giving support to the AGB scenario. Four stars appear to be single. It is very unlikely that these apparently single stars are in fact face-on systems\footnote{The probability to find one face-on orbit in their sample is $\sim$ 0.01\%.}.
These apparently single stars might have a companion with a long orbital period (about $10^3$ $-$ $10^4$ days or even longer). Nevertheless, since some CEMP-s stars are apparently single, it is worth exploring scenarios which could explain their abundances under the assumption that they are indeed single.

It has been shown that rotation at low metallicity can considerably boost the s-process in massive stars \citep{meynet06,hirschi07}. This is because of the rotational mixing operating between the He-core and H-shell during the core helium burning phase: $^{12}$C and $^{16}$O diffuse from the He-core to the H-shell, boosting the CNO cycle and forming primary $^{14}$N. When growing, the He-core engulfs the extra $^{14}$N, allowing the synthesis of extra $^{22}$Ne (via $^{14}$N($\alpha,\gamma$)$^{18}$F(,$e^+ \nu_e$)$^{18}$O($\alpha,\gamma$)$^{22}$Ne). Neutrons are then released by the reaction $^{22}$Ne($\alpha,n$). \cite{Pignatari08} did the first study of the effect of rotational mixing on the s-process at low metallicity by studying a 25 $M_{\odot}$ model with a post-processing code. \cite{frischknecht12, frischknecht16} computed massive rotating models, following the s-process during the calculation. They confirmed that rotation at low metallicity (down to [Fe/H] $\sim -3.5$) greatly enhance the s-element production in massive stars.

In this letter, we investigate whether the four apparently single CEMP-s stars might have formed from the material ejected by a massive star (source star), that lived before the birth of the CEMP-s star. 
The main difference with the AGB scenario is that the abundances of the CEMP-s stars would come from the natal cloud in which they formed. 
In the AGB scenario, only a relatively small mass fraction at the surface (received from the AGB companion) 
has the specific chemical composition making the star appear as a CEMP-s star.
We computed 14 low metallicity massive source star models with and without rotation. The s-process is followed consistently during the evolution (no post-processing). 
Models from \cite{frischknecht16} (F16 hereafter) are also considered in the analysis.
Here we discuss the models in the framework of the four apparently single CEMP-s stars. A future work will discuss in details the grid of massive stellar models with s-process and rotation. Only a few aspects of this grid are discussed here. In Sect.~\ref{sec:2} and \ref{sec:3}, we describe the computed models and compare their ejecta to the four CEMP-s stars. Conclusions are given in Sect.~\ref{sec:4}.

   \begin{figure*}
   \centering
   \begin{minipage}[c]{.49\linewidth}
     \includegraphics[scale=0.41]{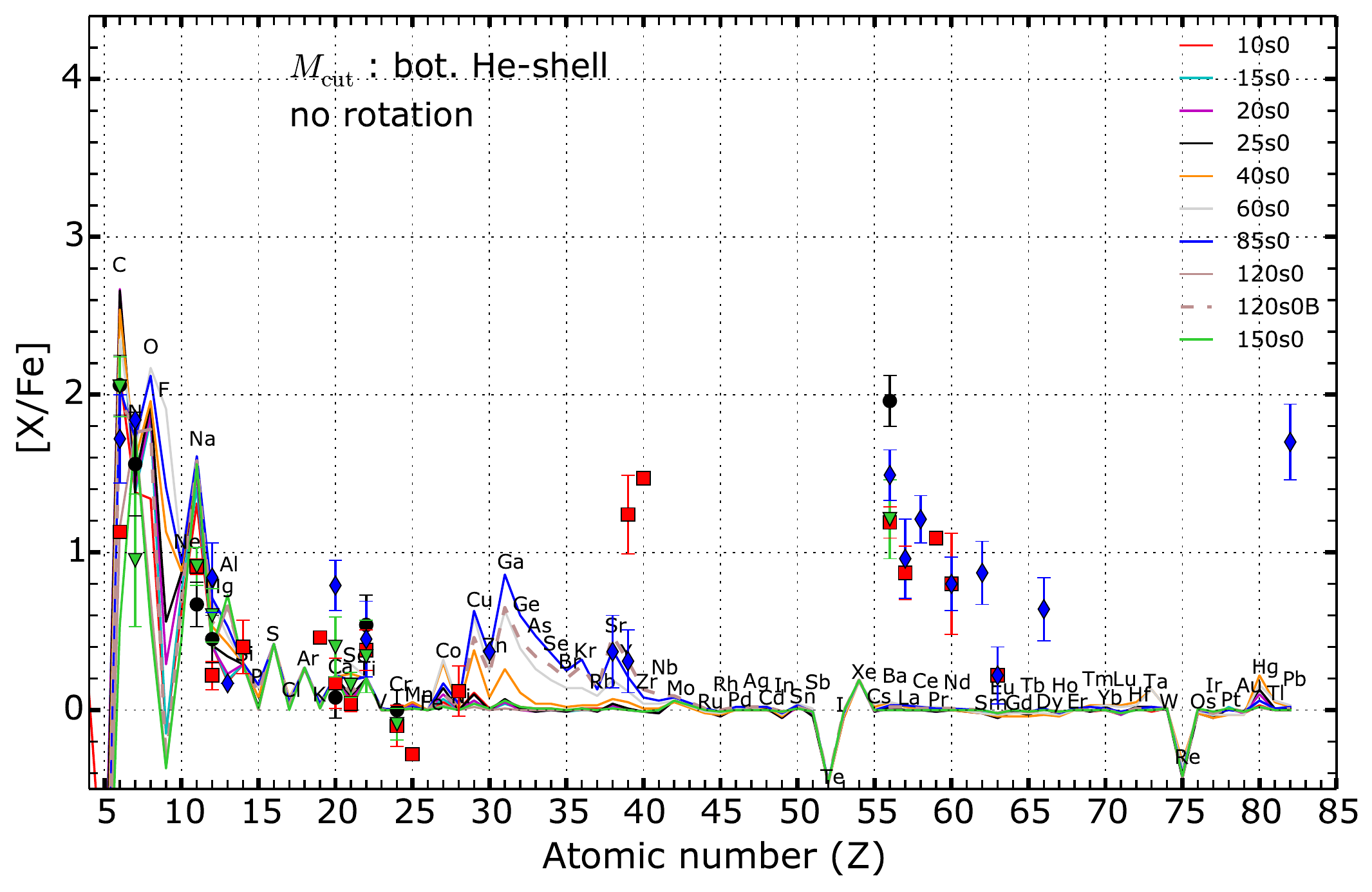}
   \end{minipage}
   \begin{minipage}[c]{.49\linewidth}
      \includegraphics[scale=0.41]{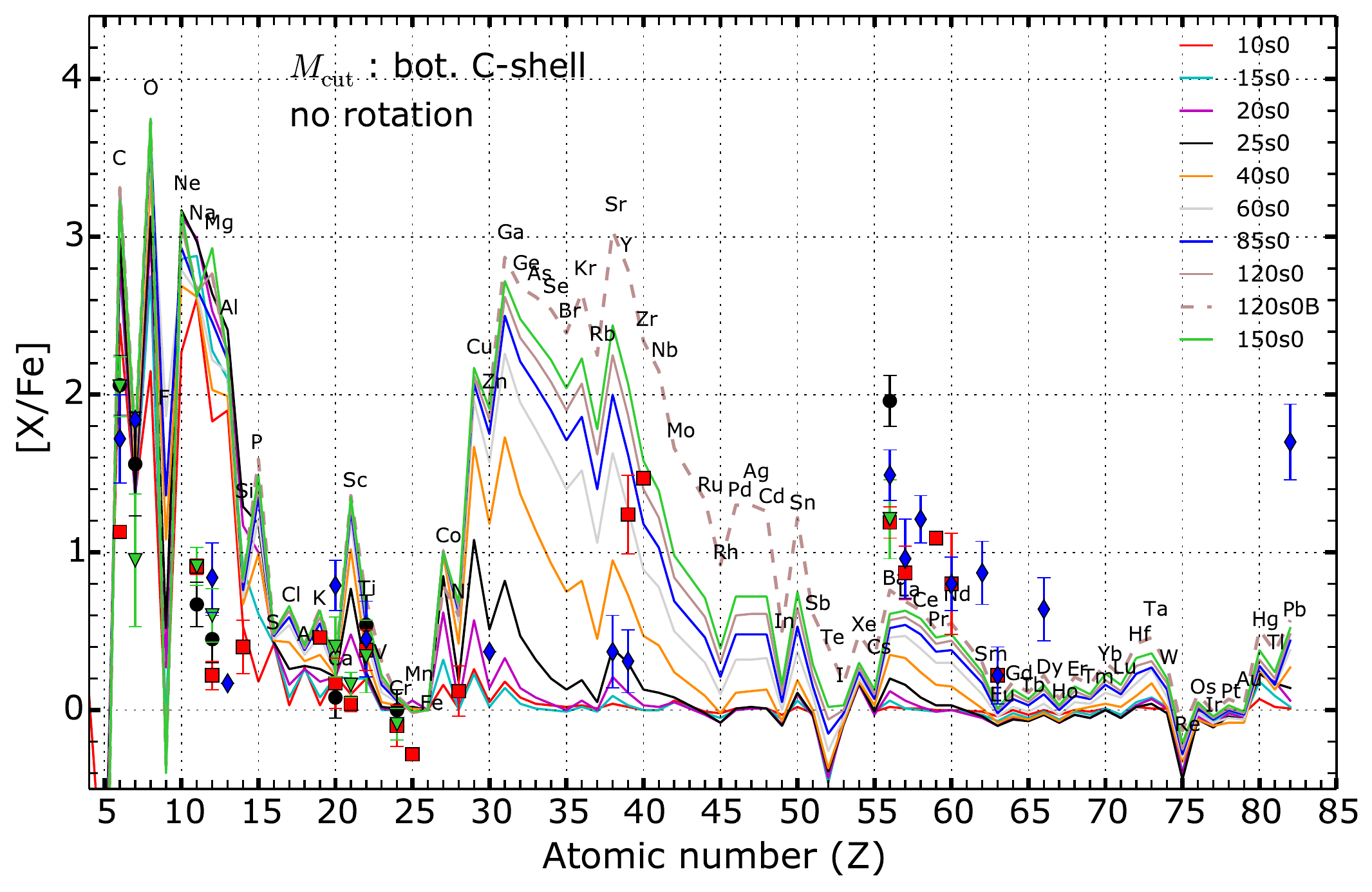}
   \end{minipage}
   \caption{Comparison of the material ejected by non-rotating source stars models with the chemical composition of four apparently single CEMP-s stars \citep{hansen16a}. 
   Different mass cuts $M_{\rm cut}$ are considered in the two panels: $M_{\rm cut}$ at the bottom of the He-shell (left panel), at the bottom of the C-shell (right panel). The models are labelled as follow: the first number is the initial mass in $M_{\odot}$, 's0' means no rotation and 'B' means that the model was computed with a lower $^{17}$O($\alpha,\gamma$) rate. The four single CEMP-s stars are shown. HE 0206-1916: black circles, HE 1045+0226: red squares, HE 2330-0555: green triangles, CS 30301-015: blue diamonds.}
\label{cemps1}
    \end{figure*}

   \begin{figure*}
   \centering
   \begin{minipage}[c]{.49\linewidth}
     \includegraphics[scale=0.41]{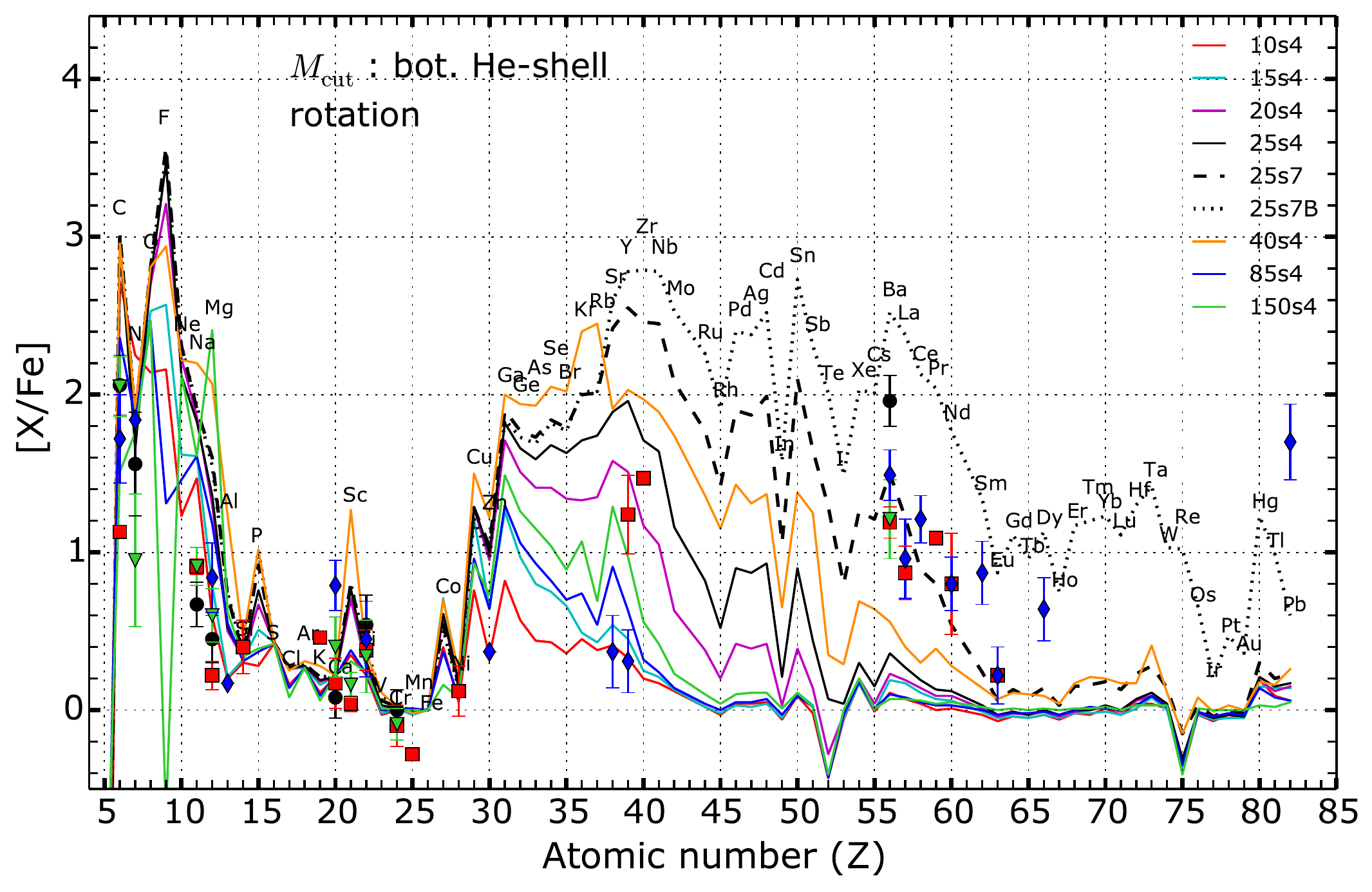}
   \end{minipage}
   \begin{minipage}[c]{.49\linewidth}
      \includegraphics[scale=0.41]{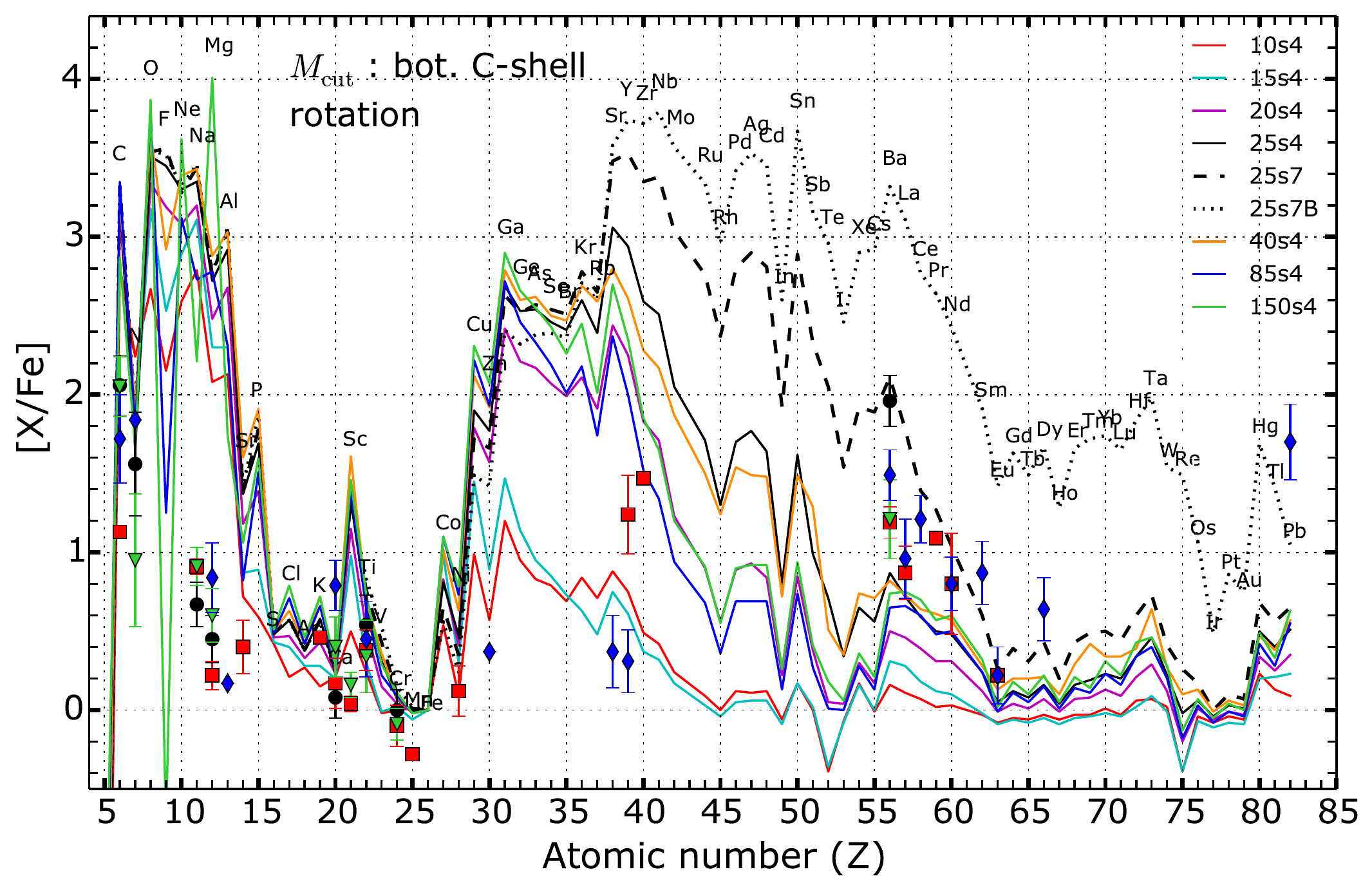}
   \end{minipage}
   \caption{Same as Fig.~\ref{cemps1} but for rotating models. In the model labels, 's4' and 's7' means rotation ($v_{\rm ini}/v_{\rm crit} = 0.4$ or $0.7$).}
\label{cemps2}
    \end{figure*}

\section{Source star models}\label{sec:2}

We use the Geneva stellar evolution code \textsc{(genec)}. \textsc{genec} is described in details \cite{eggenberger08} and \cite{ekstrom12}. 
We computed 14 rotating and non-rotating models at $Z=10^{-3}$ ([Fe/H] $= -1.8$). The initial rotation rate, $v_{\rm ini}/v_{\rm crit}$\footnote{$v_\text{crit}$ is the velocity at the equator at which the gravitational acceleration is exactly compensated by the centrifugal force.} is $0$, $0.4$ or $0.7$. Initial masses are 10, 25, 40, 60, 85, 120 and 150 $M_{\odot}$.
The nuclear network, used through all the evolution, comprises 737 isotopes, from Hydrogen to Polonium ($Z=84$). The size of the network is similar to the network of e.g. \cite{the00}, \cite{frischknecht12,frischknecht16} and follows the complete s-process.
The initial composition of metals (elements heavier than helium) is $\alpha$-enhanced (see \textsection 2.1 of F16 for more details).
Radiative mass-loss rates are from \cite{vink01} when $\log T_\text{eff} \geq 3.9$ and when $M_{\rm ini} > 15$ $M_{\odot}$. They are from \cite{jager88} if these conditions are not met.
The horizontal diffusion coefficient is from \cite{zahn92} and the shear diffusion coefficient is from \cite{talon97}.
The models are generally stopped at the end of the Ne-photodisintegration phase. In any case, the end of the C-burning phase is reached.
The s-process in massive stars mainly occur during the core He-burning phase, in the He-core. There is also a contribution from the He- and C-burning shells but that generally stays low ($\lesssim 10\%$, F16).
We used the same physical ingredients as the ones used in F16, except for some nuclear rates, which were updated. For instance, F16 used the rates of \cite{jaeger01} and \cite{angulo99} for $^{22}$Ne($\alpha,n$) and $^{22}$Ne($\alpha,\gamma$) respectively, while we used the rates of \cite{longland12}. Also, we used the new rates of \cite{best13} for $^{17}$O($\alpha,n$) and $^{17}$O($\alpha,\gamma$).
We noticed that globally, these changes have very limited effects and thus the F16 models can be consistently be used together with our models.
We investigate the impact of a $^{17}$O($\alpha,\gamma$) rate divided by 10 in two models (a 25 $M_{\odot}$ with $v_{\rm ini}$/$v_{\rm crit}$ $= 0.7$ and a 120 $M_{\odot}$ without rotation). 
A lower $^{17}$O($\alpha,\gamma$) rate favors the reaction $^{17}$O($\alpha,n$) that releases the neutrons previously captured by $^{16}$O.
Thus, the s-element production is increased. 
We tested this because the rate of $^{17}$O($\alpha,\gamma$) is still uncertain at relevant temperatures for the s-process \citep{best11} and very recent measurements tend to show that this rate is lower than expected (Laird, priv. comm.). Other rates like $^{22}$Ne($\alpha,n$) are also uncertain and can affect the results \citep{nishimura14}. They will be studied in a future work.

   \begin{figure*}
   \centering
   \begin{minipage}[c]{.49\linewidth}
      \includegraphics[scale=0.39]{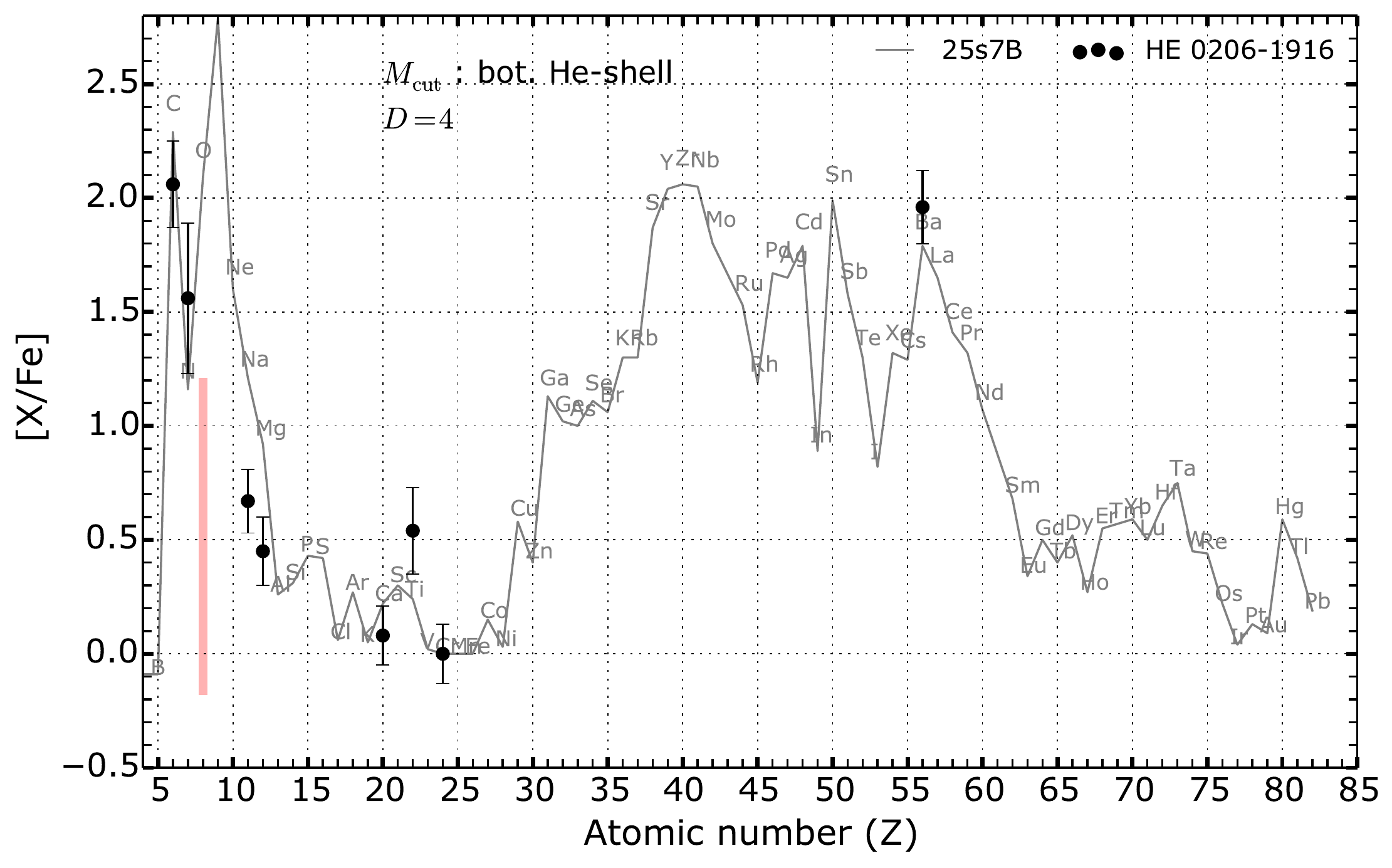}
   \end{minipage}
   \begin{minipage}[c]{.49\linewidth}
      \includegraphics[scale=0.39]{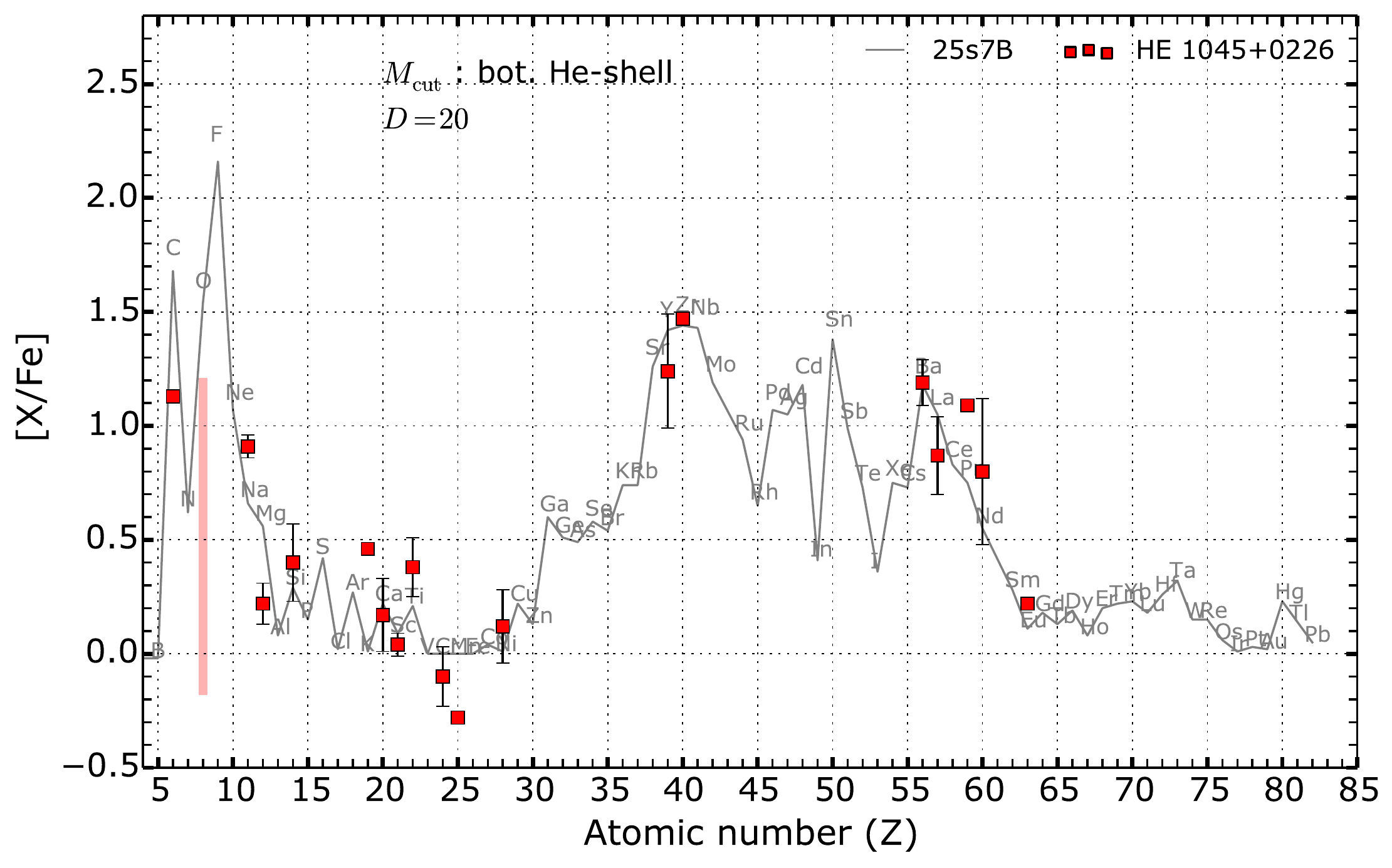}
   \end{minipage}
   \begin{minipage}[c]{.49\linewidth}
      \includegraphics[scale=0.39]{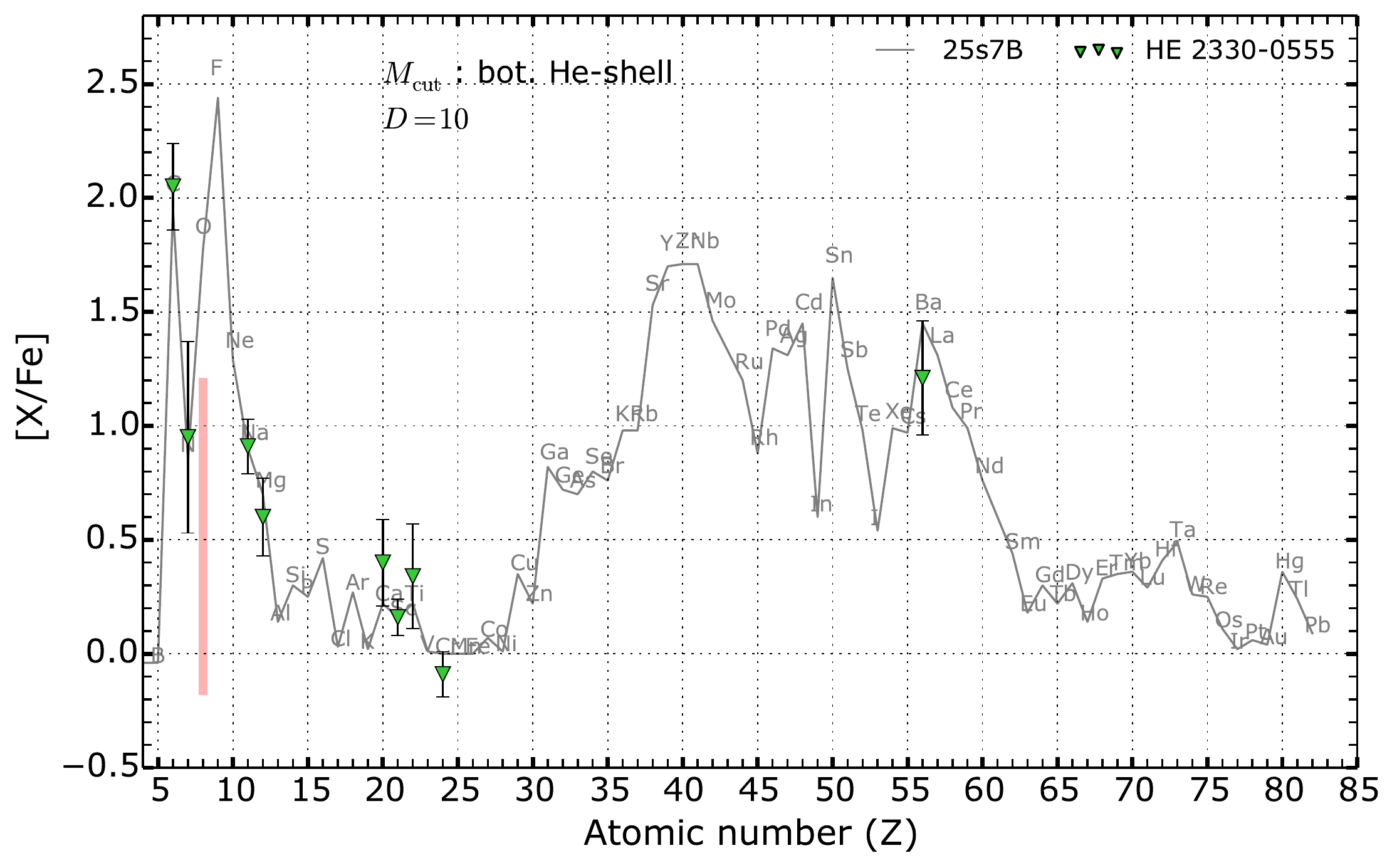}
   \end{minipage}
   \begin{minipage}[c]{.49\linewidth}
      \includegraphics[scale=0.39]{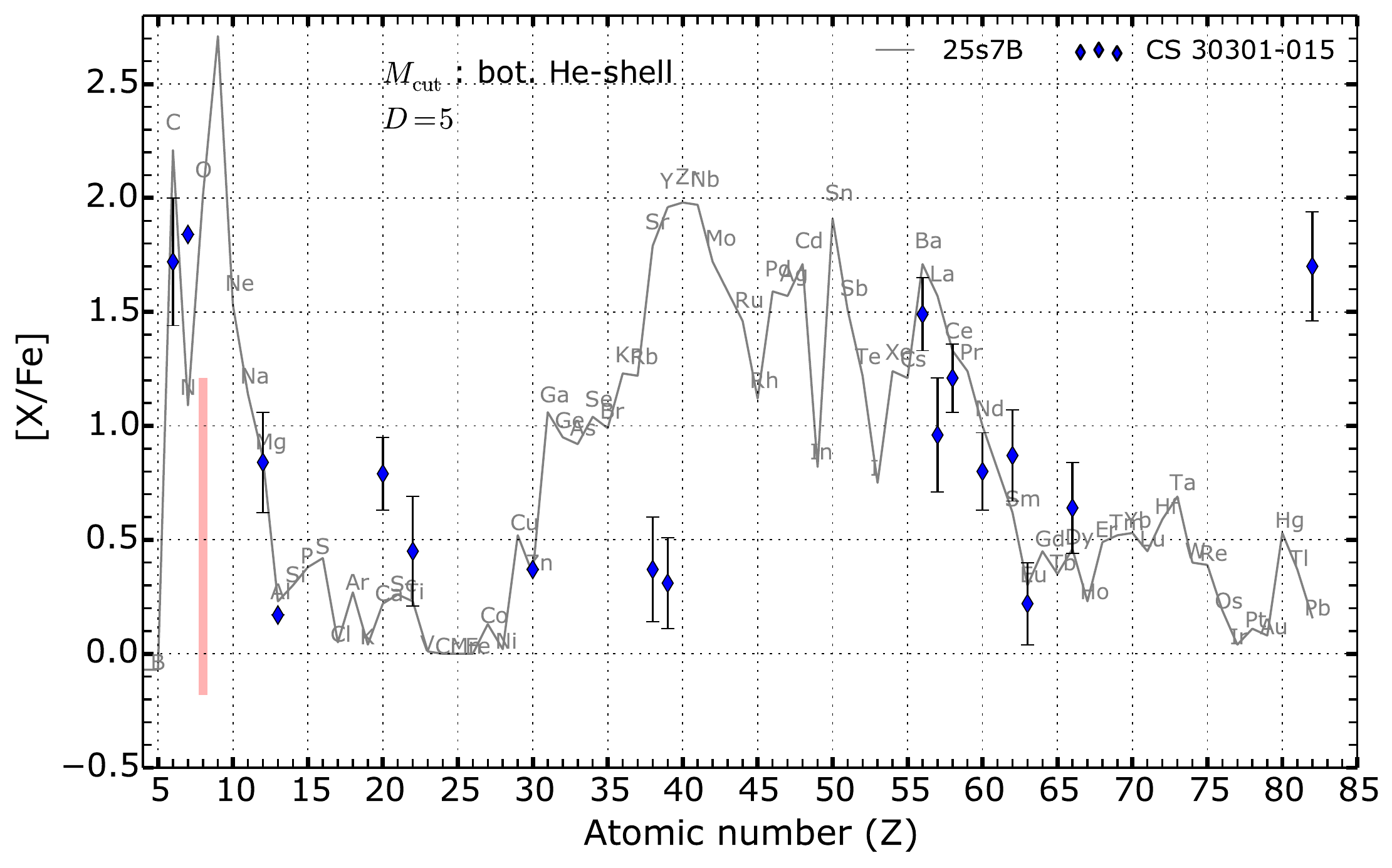}
   \end{minipage}
   \caption{Comparison of the material ejected by the 25s7B source star model (solid patterns) with the chemical composition of four apparently single CEMP-s stars \citep{hansen16a}. The ejecta of the source star is made of wind plus supernova with a mass cut set at the bottom of the He-shell. The dilution factor $D=M_{\rm ISM}/M_{\rm ej}$ is indicated. The red band at $Z=8$ shows the range of [O/Fe] ratios predicted by the AGB models of \cite{karakas10}. These models have $1<M_{\rm ini}<6$ $M_{\odot}$ and metallicities of $Z=0.004$ and $Z=0.0001$.}
\label{cemps3}
    \end{figure*}

\section{Comparison to single CEMP-s stars}\label{sec:3}

We compare the chemical composition of the material ejected by the source star models with the chemical composition observed at the surface of the four apparently single CEMP-s stars. These CEMP-s stars are HE 0206-1916 \citep{aoki07}, HE 1045+0226 \citep{cohen13}, HE 2330-0555 \citep{aoki07} and CS 30301-015 \citep{aoki02c,aoki02b}. 



Non-rotating models cannot explain the considered CEMP-s stars. They underproduce elements with $Z>55$ (See Fig.~\ref{cemps1}, the two panels correspond to different mass cuts\footnote{The mass cut delimits the part of the star which is expelled from the part which is locked into the remnant.} $M_{\rm cut}$). This is due to the secondary nature of the weak s-process in non-rotating models \citep{prantzos90}. Considering different mass cuts does not solve the problem. When expelling deep layers (Fig.~\ref{cemps1}, right panel), more s-elements are released but Na ($Z=11$) and Mg ($Z=12$) are overproduced by about 2 dex compared to the observations. Diluting the source star ejecta with the ISM will shift down the Na and Mg abundances, but also the abundances of heavier elements, with $Z>55$, which would contradict the observations.

Rotating models, especially models in the range $20-40$ $M_{\odot}$, produce significantly more s-elements than their non-rotating counterpart (see Fig.~\ref{cemps2}), due to the rotational mixing, as explained in Sect.~\ref{sec:1}. As for non-rotating models, rotating models with a deep mass cut (Fig.~\ref{cemps2}, right panel) are excluded because they overproduce Na and Mg. When considering a larger mass cut (Fig.~\ref{cemps2}, left panel), we see that only the two 25 $M_{\odot}$ models with $v_{\rm ini}/v_{\rm crit} = 0.7$ are able to produce enough elements with $Z>55$.


The chemical composition of HE 0206-1916, HE 1045+0226 and HE 2330-0555 can be reproduced by the 25 $M_{\odot}$ model with $v_{\rm ini}/v_{\rm crit} = 0.7$ and the rate of $^{17}$O($\alpha,\gamma$) divided by 10 (model '25s7B' in Fig.~\ref{cemps3}). The mass cut is set at the bottom of the He-shell and different dilution factors\footnotemark
\footnotetext{The dilution factor $D$ can be written as $D=M_{\rm ISM}/M_{\rm ej}$ with $M_{\rm ISM}$ the mass of initial ISM mixed together with $M_{\rm ej}$, the total mass ejected by the source star model (wind + supernova). For each star, $D$ was chosen in order to fit the overall observed abundance pattern as well as possible.}
are considered (see Fig.~\ref{cemps3}). There are a few discrepancies: for HE 0206-1916, Na and Mg are overestimated by $\sim 0.5$ dex, for HE 1045+0226, Na and Mg are respectively underestimated and overestimated by $\sim 0.3$ dex and C is overestimated by $\sim 0.5$ dex. It is nevertheless interesting that a single source star model with a given mass cut is able to reproduce the pattern of three CEMP-s stars. Only the dilution factor is changing.


For the last star, CS 30301-015, the trend between Ba and Dy ($56<A<66$) can be reproduced by the 25s7B model (see Fig.~\ref{cemps3}, bottom right panel).
However, the high Pb abundance ([Pb/Fe] = 1.7) cannot be explained by our models, even if considering the 25s7B model with a deep mass cut (see Fig.~\ref{cemps2}, right panel). Reducing the rate of the $^{17}$O($\alpha,\gamma$) reaction could give enough Pb, but then, Sr ($Z=38$) and Y ($Z=39$) would be even more overestimated. 
It might be that this CEMP-s star acquired its peculiar abundances owing to several sources, especially AGB stars, that are able to produce a significant amount of Pb. Further monitoring of its radial velocity might reveal a companion with a long orbital period. If not, it would mean that this CEMP-s star formed with the material ejected by one (or more) source star(s) able to synthesize enough Pb while keeping Sr ($Z=38$), Y ($Z=39$) and Eu ($Z=63$) low (Fig.~\ref{cemps3}, bottom right panel).

\section{Discussion and conclusions}\label{sec:4}

We investigated the possibility of explaining the abundances of four apparently single CEMP-s stars with the material ejected by rotating and non-rotating $10-150$ $M_{\odot}$ massive stars (source stars) that would have lived before the birth of the CEMP-s stars. 
First, we find that only layers above the bottom of the He-shell of the source star should be expelled otherwise Na and Mg abundances in the ejecta of the source stars are well above the values observed at the surface of the CEMP-s stars. Dilution with the ISM is not a solution: it will shift down the Na and Mg abundances but also the abundances of the s-elements, that would therefore be underproduced compared to the observations.
The fact that only relatively shallow layers should be expelled would be in line either with an enrichment through stellar winds only \citep{meynet06, hirschi07} and/or with a faint supernova event ending the source stellar lifetime \citep{umeda05, tominaga14}.

We find that non-rotating source star models do not provide enough s-elements with $Z>55$. 
The most favored mass range for producing the s-process elements observed at the surface of the considered CEMP-s stars is between 20 and 40 $M_{\odot}$.
An initial rotation of $v_{\rm ini}/v_{\rm crit} = 40\%$ underproduces elements with $A>55$ compared to the observations ($20-40$ $M_{\odot}$ models included). A very fast rotating 25 $M_{\odot}$ source star ($v_{\rm ini}/v_{\rm crit} = 70\%$ or $v_{\rm ini}\sim$ 500 km s$^{-1}$) gives a material able to fit the pattern of three out of the four apparently single CEMP-s stars. It is not excluded that source stars of different masses with such a high initial rotation rate could also reproduce the observed patterns.
The fourth CEMP-s star, CS 30301-015, has [Pb/Fe] $=1.7$. Such a high Pb abundance cannot be explained by our models.

Our models predict that the CEMP-s stars should have a [O/Fe] ratio about $1.5 - 2$ (see Fig.~\ref{cemps3}). Interestingly, the AGB models of \cite{karakas10} predict $-0.2<$ [O/Fe] $<1.2$ (see the red bands in Fig.~\ref{cemps3}). [O/Fe] $=1.2$ is a maximum since no dilution is assumed to obtain these values. Observing the oxygen abundance of the CEMP-s stars might therefore be a way to disentangle between the spinstar or the AGB scenario. Another way to disentangle between these scenarios would be the ls/hs ratio (ratio of light to heavy s-elements, e.g. Y/Ba). While our models predict [ls/hs] $\gtrsim 0$ \citep[see Fig.~\ref{cemps1}, \ref{cemps2} and \ref{cemps3}; also][]{chiappini11, cescutti13}, AGB models predict [ls/hs] $<0$ \citep[e.g.][]{abate15b}.

Also, our spinstar models predict $\log$($^{12}$C/$^{13}$C) $\sim 3$. HE 0206-1916 has $\log$($^{12}$C/$^{13}$C) $=1.2$ \citep{aoki07}. This discrepancy might come from the fact that HE 0206-1916 is a giant having experienced the first dredge-up: this process reduces the surface $^{12}$C/$^{13}$C ratio by bringing CNO products (mainly $^{13}$C and $^{14}$N) up to the surface. Likely, the initial surface $^{12}$C/$^{13}$C was higher. Also, a late mixing event occurring in the source star can change the CNO abundance just before the end of its evolution \citep[for details, see][we did not consider this process in the present letter]{choplin17}.

Our results suggest that fast rotating massive stars could have played a role in forming some of the CEMP-s stars. In general, our results suggest that fast rotation might have been a common phenomenon in the early universe, as already suggested by \cite{chiappini06, chiappini08} for instance.
    
\begin{acknowledgements}
We thank the anonymous referee who helped to improve this letter through constructive remarks.
This work was sponsored by the Swiss National Science Foundation (project Interacting Stars, number 200020-172505). 
RH acknowledges support from the World Premier International Research Center Initiative (WPI Initiative), MEXT, Japan and from the ChETEC COST Action (CA16117), supported by COST (European Cooperation in Science and Technology).
The research leading to these results has received funding from the European Research Council under the European Union's Seventh Framework Programme (FP/2007-2013) / ERC Grant Agreement n. 306901.
\end{acknowledgements}

\bibliographystyle{aa} 
\bibliography{biblio.bib}

\end{document}